# The vortex state (A-phase) of helimagnets $Fe_{0,5}Co_{0,5}Si$  $MnSi$, $FeGe$, as a continuous distribution of the dislocation density in the magnetic sublattice.

Braginsky Alexander Ya.

SFU, Institute of Physics, Rostov-on-don.

e-mail: **a.braginsky@mail.ru**

**Abstract.**

In this paper we propose to describe the deformed state of a helimagnet in the magnetic field (A-phase) with a pair of variables: the order parameter which characterizes the magnetic density and the tensor of distortion. Such description corresponds to discontinuities in the magnetic sublattice of the helimagnet accompanied by appearance of dislocations. It is shown that the phase diagram we found for the helimagnet is analogous to the phase diagram of the superconductors of second kind in the magnetic field, where the A-phase is a counterpart of the superconducting mixed state with the Abrikosov vortices.

.

## 1. Введение.

В последнее время появилось много экспериментальных данных подтверждающих существование вихревого состояния в магнитном поле (А-фазы) в гелимагнетиках $Fe_{0,5}Co_{0,5}Si$, $MnSi$, $FeGe$ [1-3].

Описание неоднородных магнитных структур в кристаллах развивалось как теория геликоидальных фаз [4], а в магнитном поле как киральных скирмионов [5,6]. В неравновесном потенциале в работах [5-7] учитывается вихревой инвариант $\vec{M}[\vec{\nabla}\vec{M}]$, который обобщает инвариант Лифшица для случая, когда в антисимметричной инвариантной комбинации задействовано несколько координат. Такие инварианты существуют для кристаллов, в которых нет центра инверсия, например для группы $T_4$. В [8] было показано, что наличие инварианта Лифшица приводит к фазовому переходу второго рода в неоднородное состояние.

При построении неоднородного потенциала Ландау в [6,7] полагают, что магнитный ПП представляет собой векторное поле $\vec{M}(\vec{X})$, следовательно, оно преобразуется по НП с $\vec{k}=0$. Считается, что инварианты $\vec{M}[\vec{\nabla}\vec{M}]$ описывают взаимодействие Дзялошинского-Мория $[\vec{S}_i\vec{S}_j]$ [9] в феноменологической теории Ландау.

При вычислениях в [5,6] фиксируют $|\vec{M}|=const$ из дополнительных соображений, и минимизируют неравновесный потенциал по $\vec{n}(\vec{X})=\vec{M}/|\vec{M}|$. Т.е. в [5,6] решают задачу с условием $|\vec{M}|=const$. Такая постановка задачи не учитывает уравнение $\delta\Phi/\delta|\vec{M}|=0$ в системе уравнений состояния.

С другой стороны, как известно [10], в окрестности точки фазового перехода второго рода нельзя пренебрегать уравнением $\delta\Phi/\delta|\vec{M}| = 0$. Появление новой фазы характеризуется решениями нелинейной системы уравнений состояния, для которых $|\vec{M}| \triangleright 0$, поэтому в малой окрестности точки фазового перехода второго рода необходимо решать полную систему уравнений состояния, которая содержит, в том числе, и $\delta\Phi/\delta|\vec{M}| = 0$.

В [6,7] ограничиваются квадратичным потенциалом. Как известно, приближение квадратичного потенциала справедливо в малой окрестности точки фазового перехода второго рода. При условии $|\vec{M}| = const$ нелинейное уравнение состояния $\delta\Phi/\delta|\vec{M}| = 0$ разрешимо относительно $|\vec{M}|$. Как известно, в отсутствии магнитного поля, для квадратичного приближения потенциала, всегда есть синусоидальное решение $\vec{M} = \vec{M}(\vec{X})$. Подстановка этого решения в нелинейное уравнение состояния $\delta\Phi/\delta|\vec{M}| = 0$ приводит к обнулению квадратичного члена, и как следствие, к единственному решению $|\vec{M}| = 0$ нелинейного уравнения $\delta\Phi/\delta|\vec{M}| = 0$.

Тем не менее, решения вариационной задачи с условием $|\vec{M}| = const$ в квадратичном приближении в виде киральных скирмионов в ряде случаев [6] качественно описывают неоднородные вихревые распределения магнитного момента в магнитном поле [2]. Вихревая структура уравнений задается инвариантом $\vec{M}[\vec{\nabla}\vec{M}]$, а взаимодействие $\vec{M}\vec{H}$ нарушает однородность линейного уравнения состояния, в результате чего пропадают точные периодические решения вида экспоненты.

Модель с векторным ПП $\vec{M}(\vec{X})$, не учитывает периодическую структуру гелимагнетика и не может ответить на ряд вопросов. Например, с чем связана шести лучевая дифракция в А-фазе и почему существует критическое поле, разрушающее вихревое состояние.

Существует альтернативное описание состояния А-фазы, представленное например в [11]. В [11] предлагается описывать состояния А-фазы с помощью суперпозиции шести магнитных волн плотности с $\vec{k} \neq 0$. Такое описание принципиально отличается от [6], так как изначально предполагает периодическое распределение плотности магнитного момента $\vec{M}(\vec{X})\eta_{\vec{k}_l}(\vec{X})$, где $\eta_{\vec{k}_l}(\vec{X})$ компоненты структурного ПП, который преобразуется по НП с вектором $\vec{k}_l$. По сути дела здесь модуль вектора $\vec{M}(\vec{X})$ задается структурным ПП. Компоненты $\eta_{\vec{k}_l}(\vec{X})$ характеризуют плотность распределения магнитного момента, следовательно, его можно записать как $m_{\vec{k}_l}(\vec{X}) = \vec{n}(\vec{X})\eta_{\vec{k}_l}(\vec{X})$. Такая постановка задачи с одной стороны, обобщает подход [6] (вектор $\vec{n}(\vec{X})$ характеризует направление спиновой плотности), с другой стороны, позволяет учитывать периодическую структуру распределения магнитного момента в геликоидальной фазе кристалла.

Ниже будем рассматривать НП для структурного ПП, в котором вектор $\vec{k}$ изначально не соответствует выделенной точке зоны Бриллюэна. Такое НП не эквивалентно векторному НП, например $\vec{M}(\vec{X})$ с $\vec{k}=0$, или НП с вектором $\vec{k}_0$, соответствующим выделенной точке зоны Бриллюэна, потому, что у них различное число компонент ПП, и они по разному трансформируются при преобразованиях из группы симметрии кристалла.

Выбор описания в пользу $m_{\vec{k}_l}(\vec{X}) = \vec{n}(\vec{X})\eta_{\vec{k}_l}(\vec{X})$, в качестве альтернативы векторному ПП $\vec{M}(\vec{X})$, здесь сделан не случайно. Кроме того, что он объясняет шести-лучевую дифракцию [11], такой подход в случае неоднородного распределения $\vec{k}_l(\vec{X})$ в деформированном состоянии позволяет ввести в теорию тензорное компенсирующее поле дисторсии $A_{ij}$ [12], которое приводит к вихревой структуре уравнений состояния, аналогичной уравнениям Максвелла. В этом случае неоднородное состояние гелимагнетика описывается парой переменных: ПП и тензором дисторсии. Тензор дисторсии входит как в удлиненную производную ПП, так и в потенциал в виде вихревых инвариантов $\rho_{pj} = -e_{jkn}(\partial A_{pn}/\partial X_k)$, соответствующих плотности дислокаций (здесь $e_{jkn}$ - антисимметричный единичный тензор третьего ранга). Такое описание эквивалентно описанию сверхпроводящего состояния парой переменных: сверхпроводящим ПП и электромагнитным потенциалом, в феноменологической теории Гинзбурга-Ландау.

Если обратить внимание на фазовые диаграммы гелимагнетиков в магнитном поле [1-3], то очевидна их схожесть с фазовой диаграммой для сверхпроводников второго рода [13]. Более того, состояние А-фазы в $Fe_{0,5}Co_{0,5}Si$, $MnSi$, $FeGe$ неоднократно сравнивали с вихревым состоянием Абрикосова для сверхпроводников второго рода [13].

Об аналогии описания магнитостатики и континуальной теории дислокаций известно давно [14]. В данной работе мы используем тензор дисторсии Кадича-Эделена калибровочной теории дислокаций $A_{ij}$ [14,15] в качестве компенсирующего поля структурного ПП $\eta_{\vec{k}_l}(\vec{X})$ с $k_l = \vec{k}_l(\vec{X})$. При таком описании А-фаза будет представлять собой деформацию магнитной подрешетки с образованием плотности дислокаций в каждом макроскопическом малом объеме.

## 2. Вихревые состояния гелимагнетиков с плотностью дислокаций.

Дислокации, как функции состояния, появляются в неоднородной теории Ландау, если обобщить подход Лифшица [8] для случая, когда не только величина ПП изменяется с $\vec{X}$, а также и трансформационные свойства ПП по отношению к подгруппе трансляций изменяются с макро-координатой. Действительно, трансформационные свойства ПП при трансляциях на период решетки $\vec{a}$ в общем случае характеризуются непрерывным параметром - вектором $\vec{k}$ НП: $\hat{\vec{a}}\eta_l = \exp(i\vec{k}_l\vec{a})\eta_l$, который тоже может изменяться с макро-координатой $\vec{k}_l = \vec{k}_l(\vec{X})$. В случае, когда $\vec{k}_l = \vec{k}_l(\vec{X})$ к производной $\partial \eta_l / \partial X_j$, при действии оператора трансляции $\hat{\vec{a}}$, будет добавляться величина $i\eta_l \partial(\vec{k}_l \vec{a}) / \partial X_j$, которую необходимо аннулировать при построении трансляционно-инвариантного потенциала. Для того чтобы аннулировать градиент $i\partial(\vec{k}_l \vec{a}) / \partial X_j$ в удлиненную производную ПП, по аналогии с электромагнитным потенциалом в полевой теории электродинамики, вводится дополнительное компенсирующее поле. Построение удлиненной производной проведем также как и в [16], учитывая, что локальным групповым параметром здесь является вектор $\vec{k}(\vec{X})$ НП, в отличие от скалярного группового параметра $\alpha(\vec{X})$ в электродинамике: $\hat{g}\psi = \exp(i\alpha)\psi$ [16]. Поэтому компенсирующим полем ПП с локальными трансформационными свойствами подгруппы трансляций будет тензор второго ранга $A_{pj}^l$: $D_j^l \eta_l = \left( \dfrac{\partial}{\partial X_j} - i\sum_p \kappa_p A_{pj}^l \right) \eta_l$, где $\kappa_p$ - феноменологический заряд дислокации. Очевидно, что достаточно одного тензора $A_{ij}$, чтобы компенсировать все вектора в звезде $\{\vec{k}\}$. В [12] доказано, что в качестве компенсирующего поля локального ПП можно использовать тензор дисторсии $A_{ij}$ континуальной теории дислокаций. Он задает плотность дислокаций $\rho_{pj} = -e_{jkn}(\partial A_{pn} / \partial X_k)$, по определению [17]. В этом случае неравновесный потенциал является функционалом как ПП, так и тензора дисторсии, откуда следует вихревая структура уравнений состояния $\delta\Phi / \delta A_{ij} = 0$. Градиентная инвариантность уравнений состояния в данном случае продиктована трансляционной инвариантностью локального потенциала Ландау. Так же как коэффициент перед электромагнитным потенциалом в удлиненной производной задает минимальный магнитный поток $\Phi_{\min} = \dfrac{\pi \hbar c}{e}$ [13], феноменологический заряд $\kappa_p$ задает минимальный вектор Бюргерса $\mathrm{B}_{p\min} = \dfrac{2\pi}{\kappa_p}$ [12,17].

Неоднородное распределение магнитного момента будет задаваться набором функций $\vec{m}_{\vec{k}_l}(\vec{X})$. В данном случае эти функции получаются из произведения векторного представления, характеризующего направление магнитного момента $\vec{n}(\vec{X})$ и плотности распределения магнитного момента, которая описывается компонентами структурного ПП $\eta_{\vec{k}_l}(\vec{X})$ со звездой вектора $\{\vec{k}\}$.

Вообще говоря, кристаллическая решетка обычно не взаимодействует с тензором дисторсии минимальным образом с образованием плотности дислокаций, так как это должна быть очень хрупка решетка, обусловленная слабым взаимодействием, которая также легко восстанавливается. Примером такой хрупкой упругой решетки являются жидкие кристаллы. В 1972 году де Жен обратил внимание на то, что фазовая диаграмма деформированного SmA эквивалентна фазовой диаграмме сверхпроводников второго рода [18]. Исходя из этого, он построил теорию, аналогичную теории Гинзбурга-Ландау, для SmA. В качестве основного варьируемого параметра де Жен использовал директор $\vec{n} = \vec{n}(\vec{X})$ (единичный вектор нормали к плоскости слоев смектика), так как он определяет число дислокаций: $Z_L = \oint_L \frac{\vec{n}}{d} d\vec{r}$. В [18] полагалось, что расстояние между слоями SmA не меняется $d = const$, следовательно, вектор $\vec{k}$ смектического ПП зависит от $\vec{X}$: $\vec{k}(\vec{X}) = \vec{n}(\vec{X})/d$. Т.е. можно использовать формализм [12] и удлинять производную с помощью компенсирующего тензорного поля. По сути дела де Жен впервые исследовал модель с $\vec{k} = \vec{k}(\vec{X})$.

Однако де Жен использовал векторное поле $\delta\vec{n}(\vec{X})$ в удлиненной производной, по аналогии с электромагнитным потенциалом, которое не может компенсировать изменение компонент вектора директора $\vec{n} = \vec{n}(\vec{X})$ с координатой. Кроме этого, в качестве упругой энергии в [18] использовался потенциал Франка, который содержит невихревые члены с дивергенцией: $(\vec{\nabla}\vec{n})$ [19], что не позволяло ему получить вихревые уравнения аналогичные уравнениям Лондонов. В [20] построена модель с тензорным компенсирующим полем $A_{ij}$, описывающая деформации в SmA, свободная от недостатков модели де Жена.

По аналогии со SmA, представляется, что подрешетка из магнитных спинов в гелимагнетиках, обусловленная слабым взаимодействием Дзялошинского-Мория [9], удовлетворяет двум необходимым требованиям для описания упругих свойств через тензор дисторсии: является хрупкой (деформируется с образованием дислокаций) и легко восстанавливается при снятии внешнего магнитного поля. Представляется, что гелимагнетики обладают таким свойством, так как их период не связан с кристаллической решеткой.

С одной стороны, выбор ПП в виде $m_{\vec{k}_l}(\vec{X}) = \vec{n}(\vec{X})\eta_{\vec{k}_l}(\vec{X})$ для описания гелимагнетиков в магнитном поле накладывает ограничение на интерпретацию полученных решений. Здесь предполагается существование локальной магнитной подрешетки, характеризуемой звездой вектора $\vec{k}$ в каждом макроскопическом малом объеме. С другой стороны этот выбор позволяет рассматривать деформированные состояния гелимагнетиков, в которых $k_l = \vec{k}_l(\vec{X})$, и описывать деформированное состояние тензором дисторсии.

Взаимодействие внешнего магнитного поля и магнитного ПП $\vec{n}(\vec{X})\eta_{\vec{k}}(\vec{X})$ деформирует магнитную подрешетку гелимагнетика. Разрывы магнитной подрешетки в такой модели связаны с тем, что в общем случае, нельзя построить трансляционно-инвариантную комбинацию взаимодействия $\vec{m}_{\vec{k}_l}(\vec{X})\vec{H}$. Поэтому любое взаимодействие спиновой плотности с магнитным полем приводит к разрушению геликоидальной структуры.

Модель с компенсирующим тензором дисторсии описывает структуру, которая в деформированном состоянии представляет собой гелимагнетик с непрерывным распределением дислокаций. Так как, по построению, в каждом макроскопическом малом объеме с координатой $\vec{X}$ предполагается существование локальной геликоидальной структуры, то при взаимодействии с магнитным полем, она не искривляется, а рвется по линиям дислокаций. В образовавшиеся пустоты и проникает внешнее магнитное поле. Здесь можно полагать, что внешнее магнитное поле проникает в гелимагнетик в виде квантов магнитного потока, по аналогии со смешанным состоянием сверхпроводников второго рода. В смешанном сверхпроводящем состоянии области вихрей Абрикосова характеризуются отсутствием плотности сверхпроводящих электронов. В представленном описании, в местах разрывов гелимагнитной подрешетки спиновая плотность равна нулю, по определению.

Заметим, что вихри Абрикосова также можно рассматривать как дислокации подрешетки сверхпроводящих электронов. В [12] показано, что тензор дисторсии входит в удлиненную производную сверхпроводящего ПП, и отвечает за электрон-фононное взаимодействие. Поэтому плотность дислокаций и магнитное поле сосуществуют вместе в смешанном состоянии, они задаются тензором напряжений и плотностью тока соответственно, а их потоки квантуются.

В случае $\vec{k}=\vec{k}(\vec{X})$ неоднородное распределение магнитного момента гелимагнетика приводит к появлению внутренних напряжений и дислокаций. Из калибровочной теории дислокаций Кадича-Эделена [14,15] известно об аналогии между силой Лоренца $f_i = e_{ijk} j_j B_k$ (здесь $\vec{j}$ - вектор тока, $B_k = e_{klm} \partial A_m / \partial X_l$ - магнитная индукция, $A_m$ - электромагнитный потенциал) и силой Пича-Келлера $f_i = e_{ijk} \sigma_{nj} \rho_{nk}$ (здесь $\sigma_{nj}$ - тензор напряжений). В локальной теории Ландау тензор напряжений $\sigma_{nj} = \partial \Phi / \partial A_{nj}$ зависит от компонент ПП, также как ток зависит от волновой функции в полевой теории электродинамики $j_m = \partial \Phi / \partial A_m$.

Так как отличие силы Пича-Келлера $f_i = e_{ijk} \sigma_{nj} \rho_{nk}$ от силы Лоренца $f_i = e_{ijk} j_j B_k$ по структуре состоит лишь в свертке по первому индексу тензора напряжений и плотности дислокаций, то результат взаимодействия тензора напряжений гелимагнитного ПП и плотности дислокаций имеет структуру, аналогичную вихрям Абрикосова [13].

В экспериментах, связанных с наблюдением А-фазы обычно используются тонкие пленки. Представляется что теория, описывающая вихревые магнитные состояния в тонких пленках, должна иметь решение для двумерной модели.

Вихревая модель с локальным ПП с $\vec{k}=\vec{k}(\vec{X})$ допускает плоское решение для двумерной модели и описывает А-фазу в тонких пленках. Действительно, для взаимодействия тензора напряжений и плотности дислокаций, также как для взаимодействия тока и магнитной индукции [13], есть симметричное плоское решение, которое не зависит от координаты $z$ (здесь используются общепринятые обозначения $x \equiv X_1$, $y \equiv X_2$, $z \equiv X_3$). Пусть вектор $\vec{k}$, тензор дисторсии и тензор напряжений лежат в плоскости $(xy)$: $k_n$, $A_{nj}$, $\sigma_{nj}$, где $n,j = 1,2$, тогда ненулевые компоненты плотности дислокаций имеют вид: $\rho_{n3} = \partial A_{n1}/\partial y - \partial A_{n2}/\partial x$. Следовательно, сила Пича-Келлера $f_i = e_{ij3} \sigma_{nj} \rho_{n3}$ лежит в плоскости $(xy)$, так как $i \neq 3$. В этом случае линии краевых дислокаций будут перпендикулярны плоскости $(xy)$. Таким образом, магнитная вихревая структура тонких пленок в A-фазе [2], представляет собой непрерывное распределение плотности краевых дислокаций магнитной подрешетки.

Заметим, что вихревой член $\vec{M}[\vec{\nabla}\vec{M}]$ [6] не допускает плоских решений для киральных скирмионов в тонких пленках. Действительно, пусть магнитный момент лежит в плоскости $(xy)$, тогда $\vec{M}=(M_1(x,y),M_2(x,y),0)$. Следовательно $[\vec{\nabla}\vec{M}]=(0,0,\partial M_2/\partial x - \partial M_1/\partial y)$, откуда получаем $\vec{M}[\vec{\nabla}\vec{M}]=0$.

### 3. Построение фазовой диаграммы гелимагнетиков в магнитном поле.

Прежде чем перейти к описанию фазовой диаграммы $B-T$ для гелимагнетиков надо ответить на вопрос: как связана плотность дислокаций с внешним магнитным полем. Будем исходить из того, что магнитный поток проникает в гелимагнетик вместе с плотностью дислокаций. Поток плотности дислокаций равен вектору Бюргерса, по определению. Квантование магнитного потока [13], было получено, как результат взаимодействия электромагнитного поля с волновой функцией и отражает тот факт, что электромагнитное поле компенсирует фазу волновой функции в удлиненной производной. То, что квантуется плотность дислокаций, представляется очевидным, поскольку минимальный поток плотности дислокаций равен минимальному вектору Бюргерса (минимальному периоду). На самом деле квантование плотности дислокаций указывает на то, что тензор дисторсии тоже выполняет роль компенсирующего поля в удлиненной производной ПП. В [12] показано, что в обоих случаях появление компенсирующих полей в феноменологической теории Ландау связаны с трансляционной инвариантностью локального потенциала. Электромагнитное поле связано с инвариантностью потенциала относительно временных трансляций, а тензор дисторсии с инвариантностью потенциала относительно пространственных трансляций.

В силу квантования магнитного потока, можно полагать, что существует простая пропорциональная зависимость между плотностью дислокаций и магнитным полем, которое проникает в области разрывов магнитной подрешетки гелимагнетиков. Из этого мы будем исходить при описании фазовой диаграммы гелимагнетиков $Fe_{0,5}Co_{0,5}Si$, $MnSi$, $FeGe$ в магнитном поле [1-3].

Итак, если полагать простую линейную зависимость между плотностью дислокаций и магнитным полем в неоднородном состоянии гелимагнетика, то фазовая диаграмма $Fe_{0,5}Co_{0,5}Si$ [2] в точности повторяет фазовую диаграмму сверхпроводников второго рода в магнитном поле [13].

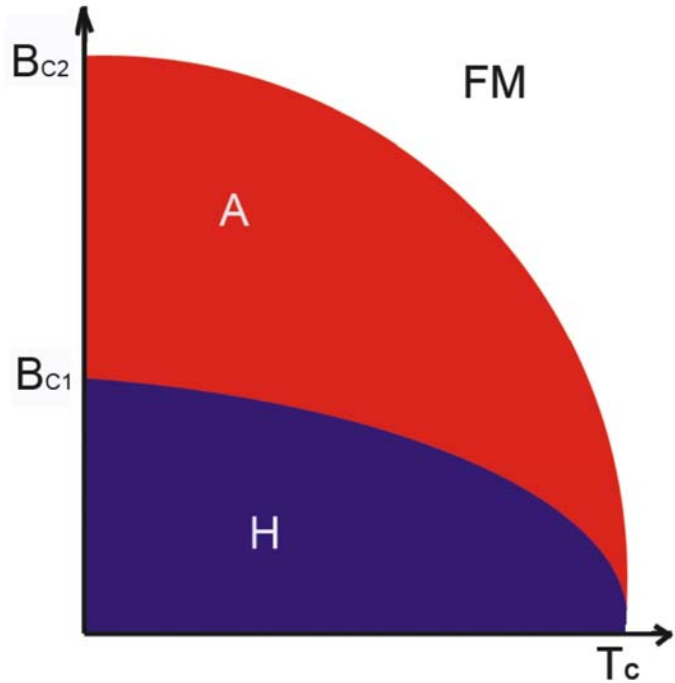

Рис.1. Фазовая диаграмма $Fe_{0,5}Co_{0,5}Si$ [2] в магнитном поле: H – гелимагнетик, A-фаза или фаза «киральных скирмионов», FM – ферромагнетик.

Действительно, из рис.1 следует аналогия между: геликоидальной фазой H и сверхпроводящей фазой; A-фазой и смешанным состоянием сверхпроводников; ферромагнитной фазой FM и несверхпроводящим состоянием [13].

Для геликоидальной фазы и сверхпроводящей фазы справедливо приближение $|\vec{M}| = const$ в магнитном поле вдали от точки фазового перехода второго рода [21]. Условие $\rho = const$, как известно [13], приводит к уравнениям Лондонов для сверхпроводящего потенциала. Уравнения состояния для магнитного ПП с удлиненной производной аналогичны уравнениям состояния Гинзбурга-Ландау [22] (если задать соответствие между плотностью дислокаций и магнитной индукцией). Поэтому существует область на фазовой диаграмме (рис.1), где геликоидальная фаза H остается стабильной, а дислокации выталкиваются на границы объема образца, подобно магнитному полю в сверхпроводнике.

При увеличении магнитного поля реализуется смешанное состояние - A-фаза (рис.1), в котором $|\vec{M}| \neq const$. В таком состоянии образуется система вихрей из дислокаций, аналогичная вихрям Абрикосова. Плотность дислокаций связана с разрывами магнитной подрешетки в неоднородном состоянии. В связи с несоразмерностью вектора $\vec{k}(\vec{X})$ магнитная подрешетка гелимагнетика не привязана к кристаллической решетке. Поэтому, в данном случае, дислокации не связаны с несовместностями кристаллической решетки, они проявляются в виде вихрей в A-фазе (рис.1).

При дальнейшем увеличении магнитного поля, можно определить критическое поле $B_{C2}$, при котором неоднородное состояние гелимагнетика разрушается и происходит переход в ферромагнитное состояние (рис.1). В этом случае, критическое магнитное поле, зависит от плотности дислокаций. Критическая плотность дислокаций $\rho_{pj} = -e_{jkn}(\partial A_{pn}/\partial X_k)$ определяется из линейного приближения уравнения состояния $\delta\Phi/\delta\eta_{\tilde{k}}$, по аналогии с определением критического магнитного поля $B_k = e_{klm}\partial A_m/\partial X_l$ из линейного приближения уравнения $\delta\Phi/\delta\psi = 0$ для сверхпроводящего ПП [13]. Критическое магнитное поле $B_{C2}$ соответствует критическому полю плотности дислокаций, при котором гелимагнитная подрешетка перестает существовать. Плотность дислокаций достигает критического значения и магнитная подрешетка разрушается.

Можно полагать, что эффект пиннинга – захвата вихрей Абрикосова дислокациями кристаллической решетки, может увеличить критическое поле и в А-фазе гелимагнетиков. С позиций термодинамики эффект пиннига может иметь простое объяснение. Плотность дислокаций магнитной подрешетки будет взаимодействовать с плотностью дислокаций кристаллической решетки, в результате чего термодинамическая энергия понизится. В данном случае плотность дислокаций кристаллической решетки будет выступать в качестве внешнего поля для плотности дислокаций гелимагнитной подрешетки. Это взаимодействие аналогично взаимодействию внешнего магнитного поля и магнитного момента.

Обратим внимание на то, что для $MnSi$ и $FeGe$ [1,3] имеют место фазовые диаграммы $B-T$, аналогичные рис.1, с той лишь разницей, что вблизи нуля температур при ненулевом магнитном поле реализуется коническая фаза с $|\vec{M}| = const$. Это согласуется с предположением Дзялошинского [21] о том, что вдали от точки фазового перехода можно использовать приближение $|\vec{M}| = const$.

В связи с появлением напряжений в гелимагнетике $\sigma_{nj} = \partial\Phi/\partial A_{nj}$, при переходе в А-фазу по температуре, должно наблюдаться упрочнение. Это упрочнение связано с наличием дислокаций в А-фазе. Оно задается дополнительными модулями упругости в потенциале, которые пропорциональны коэффициентам перед удлиненными производными ПП, поскольку тензор дисторсии входит в удлиненную производную линейным образом. Линейная зависимость тензора напряжений от тензора дисторсии, аналогична зависимости между плотностью тока и электромагнитным потенциалом в модели Гинзбурга-Ландау [13]. Выражение $\partial\sigma_{pj}/\partial A_{qi} = -K_{pjqi}$ определяет дополнительные модули упругости в А-фазе в связи с образованием неоднородного гелимагнитного упорядочения. Аналогичное упрочнение наблюдается и при переходе в сверхпроводящее состояние, оно проявляется в аномальном поведении упругих модулей при фазовом переходе второго рода [23].

**4. Заключение.**

Вихревое состояние гелимагнетиков [1-3] в магнитном поле (А-фаза) описывается тензором дисторсии, который взаимодействует с плотностью магнитного момента минимальным образом. Тензор дисторсии входит в удлиненную производную и компенсирует локальные изменения фазы плотности магнитного ПП при элементарных трансляциях. Описание неоднородного состояния гелимагнетика в магнитном поле аналогично описанию вихрей Абрикосова в теории Гинзбурга-Ландау. Сравнительный анализ рис.1 показал, что фазовая диаграмма гелимагнитиков в магнитном поле аналогична фазовой диаграмме сверхпроводников второго рода. Вихревое состояние гелимагнетика (А-фаза) является состоянием с дислокациями, которые образуются при деформации магнитной подрешетки во внешнем магнитном поле.

**Список литературы**.

[1] *S. Mühlbauer, B. Binz, F. Jonietz, C. Pfleiderer, A. Rosch, A. Neubauer, R. Georgii, P. Böni //* Science 323, 915, (2009).
[2] *X.Z. Yu, Y. Onose, N. Kanazawa, J.H. Park, J.H. Han, Y. Matsui, N. Nagaosa, Y. Tokura //* Nature 465, 901, (2010).
[3] *X. Z. Yu, N. Kanazawa, Y. Onose, K. Kimoto, W. Z. Zhang, S. Ishiwata, Y. Matsui and Y. Tokura//* Nature Materials 10, 106 (2011).
[4] *I.E. Dzyaloshinskii //* Zh. Eksp. Teor. Fiz. 46, 1420, (1964).
[5] *A.N. Bogdanov, D.A. Yablonsky //* Zh. Eksp. Teor. Fiz. 95, 178, (1989).
[6] *U.K. R¨oßler, A.N. Bogdanov, C. Pfleiderer //* Nature 442, 797 (2006).
[7] *P. Bak, H.M. Jensen, J. Phys //* C: Solid State Phys. 13 L881 (1980).
[8] *E.M. Lifshitz //* Zh. Eksp. Teor. Fiz. 11, 255 (1941).
[9] *I.E. Dzyaloshinskii //* J.Phys.Chem.Solids, 1958. Т. 4. С. 241.
[10] *L.D. Landau and E.M. Lifshitz //* Statistical Physics. Part 1, Volume 5, Nauka, Moscow 1995.
[11] *S.V. Grigoriev, et al. //* Phys. Rev., 2010. Т. B81. С. 012408.
[12] *A.Ya. Braginsky //* Фазовые переходы, упорядоченные состояния и новые материады, 2013. №.12.
[13] *E.M. Lifshitz, L.P. Pitaevskii //* Statistical Physics, Part 2. Theory of Condensed Matter, Volume 9, Nauka, Moscow, 1987.
[14] *A. Kadic, D.G. Edelen //* A Gauge Theory of Dislocations and Disclinations. Lecture Notes in Physics. Heidelberg: Springer, 1983. Т. 174.С. 168.
[15] *M. Lazar //* Mathematics and Mechanics of Solids, 2011. Т. 16. С. 253.
[16] *N.N. Bogolyubov and D.V. Shirkov //* Quantum Fields, Nauka, Moscow, 1980.
[17] *L.D. Landau and E.M. Lifshitz //* Theory of Elasticity, Volume 7, Nauka, Moscow, 1987.
[18] *P.G. De Gennes //* Solid State Commun., 1972. Т. 10. С. 753.
[19] *B.I. Halperin, T.C Lubensky //* Solid State Commun., 1974. Т. 14. С. 997.
[20] *A.Ya. Braginsky //* Phys. Rev., 2003. Т. B67. С. 174113.
[21] *I.E. Dzyaloshinskii //* Zh. Eksp. Teor. Fiz, 1964. Т. 47.С. 992.
[22] *V.L. Ginzburg, L.D. Landau //* Zh. Eksp. Teor. Fiz., 1950. Т. 20. С. 1064.
[23] *N.V. Anshukova, et al*. // Pisma v Zh. Eksp. Teor. Fiz., 1987. Т. 46. С.373.